\documentclass[english,aps,prb,twocolumn,superscriptaddress]{revtex4}
\usepackage[T1]{fontenc}
\usepackage[latin1]{inputenc}
\usepackage{amsmath}
\usepackage{graphicx}
\usepackage{amssymb}

\makeatletter



\usepackage{babel}
\makeatother

\begin{document}

\title{Unconventional pairing in the iron pnictides }

\author{Rastko Sknepnek, German Samolyuk, Yong-bin Lee, and
Jörg Schmalian}

\affiliation{Department of Physics and Astronomy and Ames Laboratory, Iowa State
University, Ames, Iowa 50011, USA }

\date{\today}

\begin{abstract}
We determine the anisotropy of the spin fluctuation induced pairing
gap on the Fermi surface of the FeAs based superconductors as function
of the exchange and Hund's coupling $J_{H}$. We find that for sufficiently
large $J_{H}$, nearly commensurate magnetic fluctuations yield a
fully gapped $s^{\pm}$-pairing state with small anisotropy of the
gap amplitude on each Fermi surface sheet, but significant variations
of the gap amplitude for different sheets of the Fermi surface. In
particular, we obtain the large variation of the gap amplitude on
different Fermi surface sheets, as seen in ARPES experiments. For
smaller values of Hund's coupling incommensurate magnetic fluctuations
yield an $s^{\pm}$-pairing state with line nodes. Such a state is
also possible once the anisotropy of the material is reduced and three
dimensional effects come into play.
\end{abstract}
\maketitle

\section{Introduction}

The recently discovered FeAs based family\cite{Kamihara08} has been
captivating the community primarily because of its high superconducting
transition temperatures, with $T_{c}$ values well above $50\mathrm{K}$
in some cases.\cite{Chen08N01,Chen08N02,Ren08N01,Wen08} While such
values for $T_{c}$ could potentially be due to the interaction between
electrons and lattice vibrations, the vibrational modes of the common
structural unit, the FeAs -planes, are rather low, making electron-phonon
interactions as the sole or primary mechanism unlikely.\cite{Boeri08}
The observation of antiferromagnetic order in undoped systems at ambient
pressure \cite{DeLaCruz08} has therefore been one of the key motivations
to explore spin fluctuations as the primary mechanism for superconductivity
in the pnictides.\cite{Mazin08,Yao08,Qui08} In this case, the role
of phonons, as intermediate boson and pairing glue, is being played
by collective paramagnon excitations of the electron fluid. In order
to determine which many body interaction is responsible for the formation
of Cooper pairs, an understanding of the symmetry and detailed momentum
dependence of the pairing gap is crucial.

Experimentally, the strongest indication that the pairing gap in the
pnictides has line nodes comes from nuclear magnetic resonance (NMR)
measurement with power law variation of the spin lattice relaxation
rate, $T_{1}^{-1}\propto T^{3}$.\cite{Nakai08,Matano08,Grafe08,Mukuda08}
On the other hand angular resolved photoemission spectroscopy (ARPES)
experiments find nodeless, weakly anisotropic gaps on the Fermi surface.\cite{Kondo08,Zhao08,Ding08}
Penetration depth measurements in the 122-compound Ba$_{0.93}$Co$_{0.07}$Fe$_{2}$As$_{2}$
support gap nodes,\cite{Gordon08} while measurements for the 1111
system NdFeAsO$_{0.9}$F$_{0.1}$ favor anisotropic gaps that remain
finite everywhere on the Fermi surface.\cite{Martin08} Interestingly,
NMR results of Ref.\cite{Matano08} and the ARPES data of Refs.\cite{Zhao08,Ding08}
are consistent to the extent that they see evidence for multiple gap
values. ARPES measurements demonstrate that the two Fermi surface
sheets around the $\Gamma$ point of Ba$_{0.6}$K$_{0.4}$Fe$_{2}$As$_{2}$
have amplitudes that differ by more than a factor $2$.\cite{Ding08}
Knight shift and spin lattice relaxation rate measurements in PrFeAsO$_{0.89}$F$_{0.11}$were
fit to two gaps with ratio $\simeq3.2$.\cite{Matano08}

In this paper we determine the momentum dependence of the superconducting
gap, where Cooper pairing is due to the exchange of antiferromagnetic
spin fluctuations. We find, in agreement with previous calculations,
Ref.\cite{Mazin08,Yao08,Cvetkovic08,Chubukov08} that the pairing
symmetry is extended $s$-wave with the gap on different Fermi surface
sheets being out of phase, i.e. we find an $s^{\pm}$ pairing state.
Superconductivity is caused by the enhanced collective spin-fluctuations
in the proximity to an ordered antiferromagnetic state. We find that
commensurate magnetic correlations can be caused by including a sufficiently
large Hund's rule coupling $J_{H}$, even in an itinerant magnetic
material. We also find that a large Hund's coupling generally yields
a stronger tendency towards superconductivity where transition temperatures
of $50\mathrm{K}$ are possible. We demonstrate that the gap function
is weakly anisotropic for most sheets of the Fermi surface, while
a significant anisotropy remains. Depending on the strength of the
exchange and Hund's couplings $J_{H}$ the gap of this Fermi surface
sheet vanishes on line nodes (for small $J_{H}$) or exhibits a moderately
anisotropic variation along the Fermi surface (for larger, more realistic
values of $J_{H}$). We also comment on the fact that a sizable interlayer
coupling, as relevant for the 122 FeAs-family, might lead to a nodal
superconducting state while for the more anisotropic 1111 family a
fully gapped state is more likely. A possible explanation for the
conflicting ARPES and NMR findings is that experiments sensitive to
the maximum of the gap, such as ARPES, see large gaps, while experiments
sensitive to the minimum of the gap, such as NMR, find node like features
due to impurity induced states in the gap.\cite{Chubukov08} The
latter is due to the fact that non-magnetic impurities in an $s^{\pm}$
pairing state behave like pair breaking magnetic impurities in a conventional
$s$-wave superconductor.

The spin fluctuation approach relies on two key assumptions:\cite{Chubukov02}
i) the proximity to a magnetic instability with paramagnons as relevant
collective modes, and ii) conventional Fermi liquid behavior away
from the instability. While electronic correlations of the Fe $3$-$d$
orbitals in the pnictides are relevant, the multi-orbital nature of
the system is likely the reason that strong local correlations, reminiscent
of a system close to a Mott insulating state do not seem to be dominating.
In addition, the carrier density of the FeAs systems does not seem
to be anywhere close to an odd number of electrons per Fe $3$-$d$
site, strongly suggesting that there are no Mott-Hubbard bands with
appreciable spectral weight. Rather, these systems are closer in their
behavior to band-insulators \ or semimetals, however with the bottom
of the electron band somewhat below the top of a hole band. The latter
leads to the observed hole and electron sheets of the Fermi surface.
Consistent with this picture is that undoped ambient pressure systems
exhibit a small but well established Drude conductivity\cite{Hu08}
and magneto-oscillations\cite{Sebastien08} in what seems to be a
partially gapped metallic antiferromagnetic state. Above the magnetic
ordering temperature a sizable Drude weight, not untypical for an
almost semimetal has been observed. The magnetic susceptibility of
BaFe$_{2}$As$_{2}$ single crystals\cite{Wang08} above the magnetic
transition is only very weakly temperature dependent and shows no
sign for local moment behavior of the Fe $3$-$d$ electron spins.
$X$-ray absorption spectroscopy for LaFeAsO$_{1-x}$F$_{x}$ is consistent
with a rigid band filling upon F-doping and moderate values for the
effective Hubbard interaction.\cite{Kroll08} Clearly, these observations
do not imply that the interactions in the FeAs systems are weak, but
rather that the phase space for strong local correlations is limited
and suggest that predominant electron-electron interactions are related
to interband scattering between the hole and electron sheets of the
Fermi surface. Despite very interesting approaches based upon the
assumption that the FeAs system are doped Mott insulators,\cite{Si08,Daghofer08}
we take the view that the iron pnictides may be good examples for
a system where collective longer ranged spin and charge excitations
play an important role. 

As shown first by Berk and Schrieffer,\cite{Berk66} magnetic fluctuations
suppress pairing for a gap function $\Delta_{a_{1}a_{2}}\left(\mathbf{p}\right)=\Delta_{0}$
that is constant as function of momentum $\mathbf{p}$ and band indices
$a_{i}$. However, changing the sign of $\Delta_{a_{1}a_{2}}\left(\mathbf{p}\right)$
as function of either $\mathbf{p}$ or $a_{1},a_{2}$ allows for nontrivial
superconducting states due to paramagnon fluctuations and makes such
fluctuations a powerful pairing mechanism. In case where only one
band contributes to the Fermi surface the sign change is a function
of momentum $\mathbf{p}$, and may lead to line or point nodes of
the gap. If there are several bands crossing the Fermi energy, strong
interband scattering can lead to a sign change of the gap between
different Fermi surface sheets, without leading to gap nodes. The
$s^{\pm}$-pairing state that results from our analysis was proposed
in the context of the $\mathrm{FeAs}$ systems in Ref.\cite{Mazin08}
in a model with structureless (in momentum state) interband pairing
interactions. In such a state, one would always obtain fully gapped
Fermi surface sheets. Our analysis shows that the model of Ref.\cite{Mazin08}
captures the $s^{\pm}$ state properly but that one needs to include
the momentum dependence of the pairing interaction to obtain states
with residual anisotropy of the pairing gap, including states that
possess nodes of the gap on a given Fermi surface sheet. A careful
investigation of the role of interband scattering in systems with
close to perfect nesting between distinct Fermi surface sheets was
performed in Refs.\cite{Cvetkovic08} and \cite{Chubukov08}. These
approaches demonstrate that under certain circumstances, pairing interactions
are enhanced due to interband nesting. At the level of the weak coupling
expansion used in Ref.\cite{Chubukov08}, this conclusion does depend
on whether the pairing mechanism is due to spin- orbital or charge
fluctuations. Our results are consistent with these findings, but
favor a spin-fluctuation mechanism boosted by intrasite, and inter-orbital
exchange and Hund's rule coupling. Our approach is closest to the
results of Refs.\cite{Kuroli08,Yao08}. The key emphasis in our work,
as compared to these interesting investigations, is to quantitatively
analyze the variation of the pairing gap on individual Fermi surface
sheets as well as between distinct sheets.

\section{The model}

Electronic structure calculations clearly show that the states close
to the Fermi level are predominantly of Fe-$3d$ character with several
sheets of the Fermi surface,\cite{singh08} as confirmed in recent
angular resolved photoemission spectroscopy (ARPES) experiments.\cite{Kondo08,Zhao08,Ding08,Liu08}
Given the need to change the sign of the gap function $\Delta_{a_{1}a_{2}}\left(\mathbf{p}\right)$,
this leads to the proposal by Mazin \emph{et al.}\cite{Mazin08}
that the gap function on sheets coupled by the magnetic wave vector
are out of phase.

We use a tight binding description of the $\mathrm{Fe}$-$d_{xz}$,
$d_{yz}$ states of the $\mathrm{FeAs}$ systems identical to the
one proposed by Raghu \emph{et al.}\cite{Raghu08} There are two
Fe atoms per crystallographic unit cell leading to the tight binding
Hamiltonian:

\begin{equation}
H_{0}=\sum_{\mathbf{p},\alpha\beta,\sigma}E_{\mathbf{p}}^{\alpha\beta}d_{\mathbf{p}\alpha\sigma}^{\dagger}d_{\mathbf{p}\beta\sigma}\end{equation}
where \ $d_{\mathbf{p}\alpha\sigma}^{\dagger}$ is the creation operator
of an electron with momentum $\mathbf{p}$ and spin $\sigma$. $\alpha$
refers to the orbital degree (i.e. $xz$ and $yz$) as well as the
label of the Fe atom within the unit cell. Momenta go from $-\pi/a$
to $\pi/a$ where $a=\sqrt{2}a_{0}$ with $\mathrm{Fe-Fe}$ distance
$a_{0}$. Thus $\widehat{E}_{\mathbf{p}}$ is a $\left(4\times4\right)$
matrix. As in Ref.\cite{Raghu08} we assume, for simplicity, that
all As atoms in the unit cell are identical. This approximation seems
justified as there are virtually no As states close to the Fermi level.
The primary relevance of the As states is only to determine the indirect
overlap between Fe orbitals on different sites. With these assumptions,
we obtain a block structure \ for the tight binding Hamiltonian of
the form\begin{equation}
\widehat{E}_{\mathbf{p}}=\widehat{h}_{\mathbf{p}}\otimes\widehat{1}+\widehat{\delta}_{\mathbf{p}}\otimes\widehat{\tau}_{x}.\end{equation}
with $\left(2\times2\right)$ unit matrix $\widehat{1}$ and Pauli
matrix $\widehat{\tau}_{x}$. $\widehat{h}_{\mathbf{p}}$ is a diagonal
$\left(2\times2\right)$ matrix with diagonal elements\begin{eqnarray}
h_{\mathbf{p}}^{11} & = & 2t_{2}\cos\left(p_{x}a\right)+2t_{3}\cos\left(p_{y}a\right)\nonumber \\
h_{\mathbf{p}}^{22} & = & 2t_{3}\cos\left(p_{x}a\right)+2t_{2}\cos\left(p_{y}a\right)\end{eqnarray}
$\ $Both diagonal elements of the $\left(2\times2\right)$ matrix
$\widehat{\delta}_{\mathbf{p}}$ are \begin{equation}
\delta_{\mathbf{p}}^{11}=\delta_{\mathbf{p}}^{22}=4t_{5}\cos\left(\frac{p_{x}a}{2}\right)\cos\left(\frac{p_{y}a}{2}\right)\end{equation}
while the off-diagonal elements are $\ $\begin{equation}
\delta_{\mathbf{p}}^{12}=\delta_{\mathbf{p}}^{21}=4t_{6}\sin\left(\frac{p_{x}a}{2}\right)\sin\left(\frac{p_{y}a}{2}\right)\end{equation}
The individual parameters, determined from fits to full potential
density functional calculations for LaFeAsO are $t_{2}=0.495\mathrm{eV}$,
$t_{3}=-0.026\mathrm{eV}$, $t_{5}=-0.026\mathrm{eV}$, $t_{6}=-0.36\,\mathrm{eV}$.

\begin{figure}

\begin{centering}
\includegraphics[width=1\columnwidth]{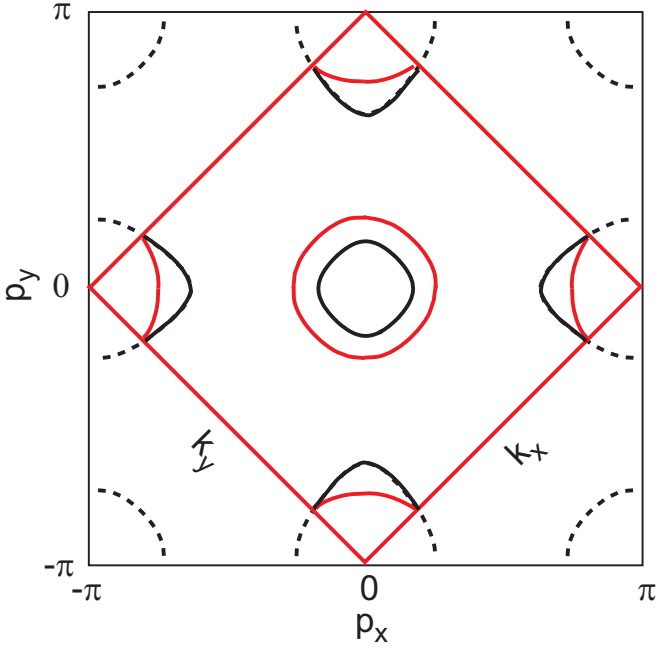}
\par\end{centering}

\caption{(Color online) Fermi surface of the tight binding parametrization
described in the text in the Brillouin zone that corresponds to two
Fe atoms per unit cell (red diamond with axes labeled by $k_{x}$
and $k_{y}$) and the larger Brillouin zone that corresponds to a
unit cell with one atom per unit cell and the larger Brillouin zone
that corresponds to a unit cell with one atom per unit cell (solid
square with axes labeled by $p_{x}$ and $p_{y}$), respectively.}

\label{fig:BZ} 
\end{figure}

Because of the assumption of treating all As atoms identically, regardless
of whether they are located above or below the Fe planes, we can describe
the system in a unit cell with only one Fe atom and can upfold the
band structure into a larger Brillouin zone, i.e. we obtain a $\left(2\times2\right)$
matrix tight binding $\widehat{\varepsilon}_{\mathbf{k}}$ in the
larger Brillouin zone. It holds $\widehat{h}_{\mathbf{p}}=\widehat{h}_{\mathbf{p+G}}$
and $\widehat{\delta}_{\mathbf{p+G}}=-\widehat{\delta}_{\mathbf{p}}$
with reciprocal lattice vector $\mathbf{G=}\left(\frac{2\pi}{a},0\right)$
and we obtain a $\widehat{\varepsilon}_{\mathbf{k}}=\widehat{h}_{\mathbf{p}}+\widehat{\delta}_{\mathbf{p}}$
for states in the original, smaller Brillouin zone and $\widehat{\varepsilon}_{\mathbf{k+G}}=\widehat{h}_{\mathbf{p}}-\widehat{\delta}_{\mathbf{p}}$
for momenta outside of it. The momentum $\mathbf{k}$ in the new,
larger Brillouin zone, with $-\frac{\pi}{a_{0}}\leq k_{x,y}<\frac{\pi}{a_{0}}$,
is given by $k_{x}=\frac{1}{\sqrt{2}}\left(p_{x}-p_{y}\right)$ and
$k_{y}=\frac{1}{\sqrt{2}}\left(p_{x}+p_{y}\right)$. For example the
wave vector of the spin density wave $\mathbf{Q=}\left(\frac{\pi}{a},\frac{\pi}{a}\right)$
becomes $\mathbf{Q=}\left(0,\frac{\pi}{a_{0}}\right)$ in the larger
BZ.

In Fig. \ref{fig:BZ} we show the Fermi surface that results from
the above tight binding parametrization at a density $n=1.05$. To
illustrate the two Brillouin zones used in the above discussion we
plot the Fermi surface in an extended zone scheme. To make contact
with Ref.\cite{Raghu08}, we note that the axis defining the $d_{xz}$
and $d_{yz}$ orbitals are rotated by $\pi/4$ relative to each other.

Next we include the local electron-electron interaction into our theory
and write \begin{eqnarray}
H_{int} & = & U\sum_{i,a}n_{ia\uparrow}n_{ia\downarrow}+U^{\prime}\sum_{i,a>b}n_{ia}n_{ib}\nonumber \\
 &  & -J_{H}\sum_{i,a>b}\left(2\mathbf{s}_{ia}\cdot\mathbf{s}_{ib}+\frac{1}{2}n_{ia}n_{ib}\right)\nonumber \\
 &  & +J\sum_{i,a>b,\sigma}d_{ia\sigma}^{\dagger}d_{ia\overline{\sigma}}^{\dagger}d_{ib\overline{\sigma}}d_{ib\sigma}\ ,\end{eqnarray}
where $n_{ia\sigma}=d_{ia\sigma}^{\dagger}d_{ia\sigma}$ is the occupation
of the orbital $a$ with spin $\sigma$ at site $i$. $n_{ia}=\sum_{\sigma}n_{ia\sigma}$
is the total charge in this orbital and $\mathbf{s}_{ia}=\frac{1}{2}\sum_{\sigma\sigma^{\prime}}d_{ia\sigma}^{\dagger}\mathbf{\sigma}_{\sigma\sigma^{\prime}}d_{ia\sigma^{\prime}}$
the corresponding spin. Thus, we include intra- and inter-orbital
direct Coulomb interactions, $U$ and $U^{\prime}$ as well as the
Hund's rule coupling $J_{H}$ and the exchange interaction $J$. The
latter are of interest as they affect the spin correlations of electrons
in different orbitals. In what follows we use $U=1\mathrm{eV}$, $U^{\prime}=0.5\mathrm{eV}$,
electron density $\rho=1.05$ per site, and we vary $J=J_{H}$ between
$J=0$ and $J=0.5\mathrm{eV}$ to explore the role of the exchange
and Hund's interactions on the pairing state. Recent $X$-ray absorption
spectroscopy measurements support values for the Hund's couplings
that lead to a preferred high spin configuration,\cite{Kroll08}
leading to larger values of $J_{H}\lesssim U$.

The interaction term can be put into a more compact form\cite{multiorbitals}\begin{equation}
H_{int}=\frac{1}{4}\sum_{i,a_{l};\sigma_{l}}U_{\sigma_{1}\sigma_{2},\sigma_{3}\sigma_{4}}^{a_{1}a_{2},a_{3}a_{4}}d_{ia_{1}\sigma_{1}}^{\dagger}d_{ia_{2}\sigma_{2}}^{\dagger}d_{ia_{3}\sigma_{3}}d_{ia_{4}\sigma_{4}}\end{equation}
and, in the absence of spin orbit interaction, split into a spin and
a charge contribution:\begin{eqnarray}
U_{\sigma_{1}\sigma_{2},\sigma_{3}\sigma_{4}}^{a_{1}a_{2},a_{3}a_{4}} & = & -\frac{1}{2}U_{s}^{a_{1}a_{4},a_{2}a_{3}}\mathbf{\sigma}_{\sigma_{1}\sigma_{4}}\cdot\mathbf{\sigma}_{\sigma_{2}\sigma_{3}}\nonumber \\
 &  & +\frac{1}{2}U_{c}^{a_{1}a_{4},a_{2}a_{3}}\delta_{\sigma_{1}\sigma_{4}}\delta_{\sigma_{2}\sigma_{3}}.\end{eqnarray}
The above Hamiltonian is then recovered if we chose \begin{equation}
U_{s}^{a_{1}a_{4},a_{2}a_{3}}=\left\{ \begin{array}{cc}
U & \text{if \ }a_{1}=a_{2}=a_{3}=a_{4}\\
U^{\prime} & \text{if \ }a_{1}=a_{3}\neq a_{2}=a_{4}\\
J_{H} & \text{if \ }a_{1}=a_{4}\neq a_{2}=a_{3}\\
J & \text{if \ }a_{1}=a_{2}\neq a_{3}=a_{4}\end{array}\right.\end{equation}
for the spin part of the interaction, and \begin{equation}
U_{c}^{a_{1}a_{4},a_{2}a_{3}}=\left\{ \begin{array}{cc}
U & \text{if \ }a_{1}=a_{2}=a_{3}=a_{4}\\
-U^{\prime}+2J_{H} & \text{if \ }a_{1}=a_{3}\neq a_{2}=a_{4}\\
2U^{\prime}-J_{H} & \text{if \ }a_{1}=a_{4}\neq a_{2}=a_{3}\\
J & \text{if \ }a_{1}=a_{2}\neq a_{3}=a_{4}\end{array}\right..\end{equation}
for the corresponding charge contribution, respectively.

\subsection{Collective spin and charge fluctuations}

We determine the single particle and collective magnetic excitation
spectrum within a self consistent one loop approach, the multiple
orbital version\cite{multiorbitals,schmalian} of the fluctuation
exchange approximation of Ref.\cite{Bickers89}. Once we have self
consistently determined the fermionic Green's function $G^{ab}\left(k\right)$
where $k=\left(\mathbf{k},\omega_{n}\right)$ stands jointly for the
crystal momentum $\mathbf{k}$ and the Matsubara frequency $\omega_{n}=\left(2n+1\right)\pi T$,
we determine the symmetry of the pairing state from the linearized
gap equation. In the normal state, the matrix Green's function of
the problem is \begin{equation}
\widehat{G}\left(k\right)=\left(i\omega_{n}\widehat{1}-\widehat{\varepsilon}_{\mathbf{k}}-\widehat{\Sigma}\left(k\right)\right)^{-1}\end{equation}
where $\widehat{G}_{k}$, $\widehat{\Sigma}_{k}$, $\widehat{\varepsilon}_{\mathbf{k}}$
are all $2\times2$ matrices in orbital space in the larger Brillouin
zone. The self energy is given as a sum of a Hartree-Fock contribution
and a fluctuation term\begin{equation}
\Sigma^{a_{1}a_{2}}\left(k\right)=\sum_{k^{\prime}}\sum_{a_{3}a_{4}}G^{a_{3}a_{4}}\left(k^{\prime}\right)\Gamma_{ph}^{a_{1}a_{3},a_{4}a_{2}}\left(k-k^{\prime}\right)\label{ns1}\end{equation}
where $\sum_{k}\ldots=\frac{T}{N^{2}}\sum_{\mathbf{k},n}\ldots$ includes
the summation over momenta and over Matsubara frequencies.

Introducing the particle quantum numbers $A=\left(a_{1},a_{2}\right)$
and $B=\left(a_{3},a_{4}\right)$ labeling the rows and columns of
two particle states interaction, $\Gamma_{ph}^{a_{1}a_{3},a_{4}a_{2}}\left(q\right)=\Gamma_{ph}^{AB}\left(q\right)$
becomes a $4\times4$-dimensional symmetric operator $\widetilde{\Gamma}_{ph}\left(q\right)$.
Similarly we obtain in this two particle basis a matrix representation
for the spin and charge couplings $\widetilde{U}^{s}$ and $\widetilde{U}^{c}$:
\begin{equation}
\widetilde{U}^{s}=\left(\begin{array}{cccc}
U & 0 & 0 & J_{H}\\
0 & U^{\prime} & J & 0\\
0 & J & U^{\prime} & 0\\
J_{H} & 0 & 0 & U\end{array}\right)\end{equation}
and\begin{equation}
\widetilde{U}^{c}=\left(\begin{array}{cccc}
U & 0 & 0 & W\\
0 & W^{\prime} & J^{\prime} & 0\\
0 & J^{\prime} & W^{\prime} & 0\\
W & 0 & 0 & U\end{array}\right)\end{equation}
where $W=2U^{\prime}-J_{H}$ and $W^{\prime}=2J_{H}-U^{\prime}$.
In this two particle formalism it is now straightforward to sum particle-hole
ladder and bubble diagrams and it follows \begin{equation}
\widetilde{\Gamma}_{ph}\left(q\right)=\frac{3}{2}\widetilde{V}^{s}\left(q\right)+\frac{1}{2}\widetilde{V}^{s}\left(q\right)\label{ns2}\end{equation}
with \begin{eqnarray}
\widetilde{V}^{s}\left(q\right) & = & \widetilde{U}^{s}\left(1-\widetilde{\chi}\left(q\right)\widetilde{U}^{s}\right)^{-1}\widetilde{\chi}\left(q\right)\widetilde{U}^{s}\label{ns3}\\
 &  & -\frac{1}{2}\widetilde{U}^{s}\widetilde{\chi}\left(q\right)\widetilde{U}^{s}\nonumber \\
\widetilde{V}^{c}\left(q\right) & = & \widetilde{U}^{c}\left(1+\widetilde{\chi}\left(q\right)\widetilde{U}^{c}\right)^{-1}\widetilde{\chi}\left(q\right)\widetilde{U}^{c}\nonumber \\
 &  & -\frac{1}{2}\widetilde{U}^{c}\widetilde{\chi}\left(q\right)\widetilde{U}^{c}\ .\end{eqnarray}
Here $\widetilde{\chi}\left(q\right)$ is the matrix of particle-hole
bubble in the two particle basis. Explicitly it holds: \begin{equation}
\chi^{a_{1}a_{2},a_{3}a_{4}}\left(q\right)=-\frac{T}{N^{2}}\sum_{k}G^{a_{2}a_{3}}\left(k+q\right)G^{a_{4}a_{1}}\left(q\right).\label{chidef}\end{equation}
The Hartree-Fock term of the self energy \begin{equation}
\Sigma_{HF}^{a_{1}a_{2}}=\sum_{a_{3}a_{4}}\left(\frac{3}{2}\widetilde{U}^{s,a_{3}a_{1},a_{4}a_{2}}-\frac{1}{2}\widetilde{U}^{c,a_{3}a_{1},a_{4}a_{2}}\right)G_{\mathbf{0}}^{a_{3}a_{4}}\left(\tau^{-}\right)\end{equation}
is frequency and momentum independent and determined by $G_{\mathbf{0}}^{a_{3}a_{4}}\left(\tau^{-}\right)=\left\langle d_{\mathbf{0}a_{3}}^{\dagger}d_{\mathbf{0}a_{4}}\right\rangle $.
It holds for the diagonal elements \begin{eqnarray}
\Sigma_{HF}^{a_{1}a_{1}} & = & Un_{a_{1}}+\left(2U^{\prime}-J_{H}\right)\sum_{a_{2}\neq a_{1}}n_{a_{2}}\nonumber \\
 & = & \left(U-2U^{\prime}+J_{H}\right)n_{a_{1}}+\left(2U^{\prime}-J_{H}\right)n\end{eqnarray}
whereas the off-diagonal elements ($a_{1}\neq a_{2}$) are given as:
\begin{equation}
\Sigma_{HF}^{a_{1}a_{2}}=\left(2J_{H}+J-U^{\prime}\right)\left\langle d_{\mathbf{0,}a_{1}}^{\dagger}d_{\mathbf{0,}a_{2}}\right\rangle .\end{equation}

We are interested in the superconducting transition temperature and
the symmetry of the superconducting state, determined by the corresponding
anomalous self energy $\widehat{\Phi}_{\mathbf{k}}\left(\omega_{n}\right)$.
Summing up the same class of diagrams in the superconducting state
yields\begin{equation}
\Phi^{a_{1}a_{2}}\left(k\right)=\sum_{k^{\prime}a_{3}a_{4}}\Gamma_{pp}^{a_{3}a_{1},a_{2}a_{4}}\left(k-k^{\prime}\right)F^{a_{3}a_{4}}\left(k^{\prime}\right),\label{gap1}\end{equation}
with Gor'kov function $\widehat{F}\left(k\right)$. $\widetilde{\Gamma}_{pp}\left(q\right)$
is the corresponding operator in the two particle representation.
In this paper we only solve the linearized version of Eq. (\ref{gap1})
to determine the superconducting transition temperature as well as
the nature of the pairing state right below $T_{c}$. Close to the
superconducting transition temperature we linearize the anomalous
propagator \begin{equation}
\widehat{F}\left(k\right)\simeq-\widehat{G}\left(k\right)\widehat{\Phi}\left(k\right)\widehat{G}\left(-k\right)\end{equation}
and obtain\begin{eqnarray}
\Phi^{a_{1}a_{2}}\left(k\right) & = & -\frac{T}{N^{2}}\sum_{k^{\prime}a_{3}a_{4}a_{5}a_{6}}\Gamma_{pp}^{a_{3}a_{1},a_{2}a_{4}}\left(k-k^{\prime}\right)\nonumber \\
 &  & G^{a_{3}a_{5}}\left(k^{\prime}\right)\Phi^{a_{5}a_{6}}\left(k^{\prime}\right)G^{a_{6}a_{4}}\left(-k^{\prime}\right).\label{lingap}\end{eqnarray}
Since $F^{a_{3}a_{4}}\left(k^{\prime}\right)$ is of first order in
the anomalous self energy $\Phi^{a_{1}a_{2}}\left(k\right)$, the
linearized version of Eq. (\ref{lingap}) is determined by $\widetilde{\Gamma}_{pp}\left(q\right)$
for $\Phi^{a_{1}a_{2}}\left(k\right)=0$. In this limit it follows,
after summing the same bubble and ladder diagrams as for $\widetilde{\Gamma}_{ph}\left(q\right)$
that \begin{equation}
\widetilde{\Gamma}_{pp}\left(q\right)=\frac{3}{2}\widetilde{V}^{s}\left(q\right)-\frac{1}{2}\widetilde{V}^{c}\left(q\right)\end{equation}
with \begin{eqnarray}
\widetilde{V}^{s}\left(q\right) & = & \widetilde{U}^{s}\left(1-\widetilde{\chi}\left(q\right)\widetilde{U}^{s}\right)^{-1}\widetilde{\chi}\left(q\right)\widetilde{U}^{s}+\frac{\widetilde{U}^{s}}{2}\nonumber \\
\widetilde{V}^{c}\left(q\right) & = & \widetilde{U}^{c}\left(1+\widetilde{\chi}\left(q\right)\widetilde{U}^{c}\right)^{-1}\widetilde{\chi}\left(q\right)\widetilde{U}^{c}\ -\frac{\widetilde{U}^{c}}{2}.\end{eqnarray}

In what follows we first solve the coupled equations Eqs. (\ref{ns1}),
(\ref{ns2}), (\ref{ns3}) and (\ref{chidef}) in the normal state
on a $32\times32$ lattice with $2^{11}$ Matsubara frequencies. The
solutions of the normal state equations are then used to solve the
linearized equation for the superconducting self energy. In order
to determine the superconducting transition temperature we replace
$\Phi^{a_{1}a_{2}}\left(p\right)$ on the l.h.s. of Eq. (\ref{lingap})
by $\lambda\Phi^{a_{1}a_{2}}\left(p\right)$. The resulting eigenvalue
equation yields an eigenvalue $\lambda=1$ if $T=T_{c}$, i.e. the
temperature where the linearization is permitted. For $T>T_{c}$,
it holds $\lambda<1$ for the largest eigenvalue. Even if $\lambda<1$,
the result is still useful as $\left(1-\lambda\right)^{-1}$ is proportional
to the pairing correlation function. Most importantly, the eigenvector
of the leading eigenvalue determines the momentum and band-index dependence
of the gap right below $T_{c}$. In order to simplify the above eigenvalue
equation we replace $\widetilde{\Gamma}_{pp}\left(p\right)$ by its
zero Matsubara frequency value, i.e., $\widetilde{\Gamma}_{pp}\left(\mathbf{p},\omega_{n}=0\right)$.
Thus, we keep the dynamic excitations that determine the frequency
dependence of the normal state single particle self energy, but assume
that the dynamics of the pairing interaction is structureless. Such
an approximation would be problematic close to a magnetic quantum
critical point with diverging antiferromagnetic correlation length,\cite{Chubukov05}
but is expected to be reasonable for intermediate magnetic correlations,
as seems to be the case in the FeAs systems. A consequence of this
approximation is that we lose the information about the frequency
dependence of the anomalous self energy. We keep its momentum and
orbital index dependence.

\subsection{Symmetry considerations}

For a proper interpretation of the momentum dependence of the superconducting
gap in a multi orbital problem, we analyze the point group symmetry
of the two band model describing the $d_{xz}$ and $d_{yz}$ orbitals.
We consider the behavior of the Hamiltonian under the tetragonal point
group $D_{4h}=C_{4v}\otimes C_{i}$ where $C_{i}$ is the inversion
and $C_{4v}$ contains next to the identity $E$ two four-fold rotations
$c_{4}$ one two-fold rotation $c_{2}$, two mirror reflexions along
the axis $\sigma_{v}$ and two mirror reflexions along the diagonals
$\sigma_{d}$. The Hamiltonian is invariant with respect to the group
$D_{4h}$, i.e. \begin{equation}
\widehat{\varepsilon}_{\mathbf{k}}=R\widehat{\varepsilon}_{\mathbf{k}}\text{ for all }R\in D_{4h}\end{equation}
Since the two orbitals $d_{xz}$ and $d_{yz}$ transform like coordinates
for in-plane symmetry operations, it holds \begin{equation}
R\widehat{\varepsilon}_{\mathbf{k}}=D_{R}^{\left(1\right)}\widehat{\varepsilon}_{D_{R}^{\left(1\right)}\mathbf{k}}\left(D_{R}^{\left(1\right)}\right)^{-1}\end{equation}
where $D_{R}^{\left(1\right)}$ is the representation of $R$ which
transforms the coordinates. It then follows that the spinor \begin{equation}
c_{\mathbf{k}\sigma}=\left(\begin{array}{c}
c_{\mathbf{k,}xz,\sigma}\\
c_{\mathbf{k,}yz},\sigma\end{array}\right)\end{equation}
transforms as \begin{equation}
Rc_{\mathbf{k}}=D_{R}^{\left(1\right)-1}c_{D_{R}^{\left(1\right)}\mathbf{k}}\end{equation}
which determines the transformation properties of the superconducting
gap function in the singlet channel: \begin{equation}
R\Phi_{\mathbf{k}}^{ab}=\sum_{a^{\prime}b^{\prime}}D_{Raa^{\prime}}^{\left(1\right)-1}D_{Rbb^{\prime}}^{\left(1\right)-1}\Phi_{D_{R}^{\left(1\right)}\mathbf{k}}^{a^{\prime}b^{\prime}}\ .\end{equation}
It follows for the transformation of the gap under the point group
operations: \begin{eqnarray}
E\widehat{\Phi}_{\left(k_{x},k_{y}\right)} & = & \left(\begin{array}{cc}
\Phi_{\left(k_{x},k_{y}\right)}^{xx} & \Phi_{\left(k_{x},k_{y}\right)}^{xy}\\
\Phi_{\left(k_{x},k_{y}\right)}^{yx} & \Phi_{\left(k_{x},k_{y}\right)}^{yy}\end{array}\right)\nonumber \\
C_{4}\widehat{\Phi}_{\left(k_{x},k_{y}\right)} & = & \left(\begin{array}{cc}
\Phi_{\left(k_{y},-k_{x}\right)}^{yy} & -\Phi_{\left(k_{y},-k_{x}\right)}^{yx}\\
-\Phi_{\left(k_{y},-k_{x}\right)}^{xy} & \Phi_{\left(k_{y},-k_{x}\right)}^{xx}\end{array}\right)\nonumber \\
C_{2}\widehat{\Phi}_{\left(k_{x},k_{y}\right)} & = & \left(\begin{array}{cc}
\Phi_{\left(-k_{x},-k_{y}\right)}^{xx} & \Phi_{\left(-k_{x},-k_{y}\right)}^{xy}\\
\Phi_{\left(-k_{x},-k_{y}\right)}^{yx} & \Phi_{\left(-k_{x},-k_{y}\right)}^{yy}\end{array}\right)\nonumber \\
\sigma_{v}\widehat{\Phi}_{\left(k_{x},k_{y}\right)} & = & \left(\begin{array}{cc}
\Phi_{\left(k_{x},-k_{y}\right)}^{xx} & -\Phi_{\left(k_{x},-k_{y}\right)}^{xy}\\
-\Phi_{\left(k_{x},-k_{y}\right)}^{yy} & \Phi_{\left(k_{x},-k_{y}\right)}^{yy}\end{array}\right)\nonumber \\
\sigma_{d}\widehat{\Phi}_{\left(k_{x},k_{y}\right)} & = & \left(\begin{array}{cc}
\Phi_{\left(k_{y},k_{x}\right)}^{yy} & \Phi_{\left(k_{y},k_{x}\right)}^{yx}\\
\Phi_{\left(k_{y},k_{x}\right)}^{xy} & \Phi_{\left(k_{y},k_{x}\right)}^{xx}\end{array}\right)\end{eqnarray}
A rotation by $\pi/2$ that causes a sign change of an off-diagonal
element of $\widehat{\Phi}$ is therefore no indication for pairing
in the d-wave channel. Thus assuming a rotation by $\pi/2~$(as generated
by $C_{4}$) yields a sign change of the off-diagonal element and
no such change occurs for the diagonal element, we find $C_{4}\widehat{\Phi}_{\left(k_{x},k_{y}\right)}=\widehat{\Phi}_{\left(k_{x},k_{y}\right)}$,
i.e. the gap belongs either to the irreducible representation $A_{1}$
or $A_{2}$. If furthermore the gap doesn't change sign upon reflection
on the axis we conclude it is $A_{1}$, corresponding to $s$-wave
pairing. This will be the case in our subsequent analysis of the numerical
solution of spin fluctuation induced pairing.

\subsection{Results}

\begin{figure}
\begin{centering}
\includegraphics[width=0.97\columnwidth]{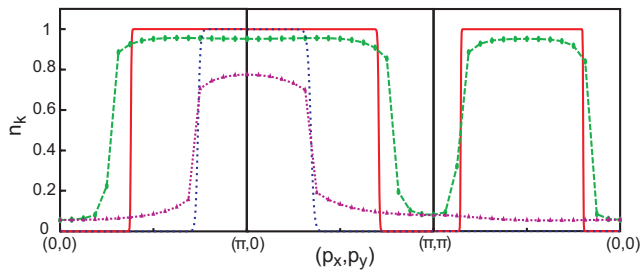}
\par\end{centering}

\caption{(color online) Band occupation number along $\left(0,0\right)\to\left(\pi,0\right)\to\left(\pi,\pi\right)\to\left(0,0\right)$
for noninteracting case (red and blue) and with interactions (green
and violet). }
\label{fig:occupation}
\end{figure}

In Fig. \ref{fig:occupation} we show the occupation number $n_{\mathbf{p}}$
determined from the full solution of the self consistent equations
in the normal state at $T=0.004\mathrm{eV}$. We compare our results
with the corresponding occupation of the tight binding model without
interaction at the same filling. The electron band closest to $\mathbf{p=}\left(\pi/a,0\right)$
undergoes a substantial distribution of carriers as it is being pushed
very close to the Fermi energy. Similarly we observe a decrease in
the Fermi surface volume of the hole band centered around $\mathbf{p=}\left(0,0\right)$.
Still the overall shape and topology of the various Fermi surface
sheets are unchanged by many body interactions.

\begin{figure}
\begin{minipage}[t][1\totalheight]{1\columnwidth}%
\begin{flushleft}
a)
\par\end{flushleft}

\begin{center}
\includegraphics[angle=270,width=1\columnwidth]{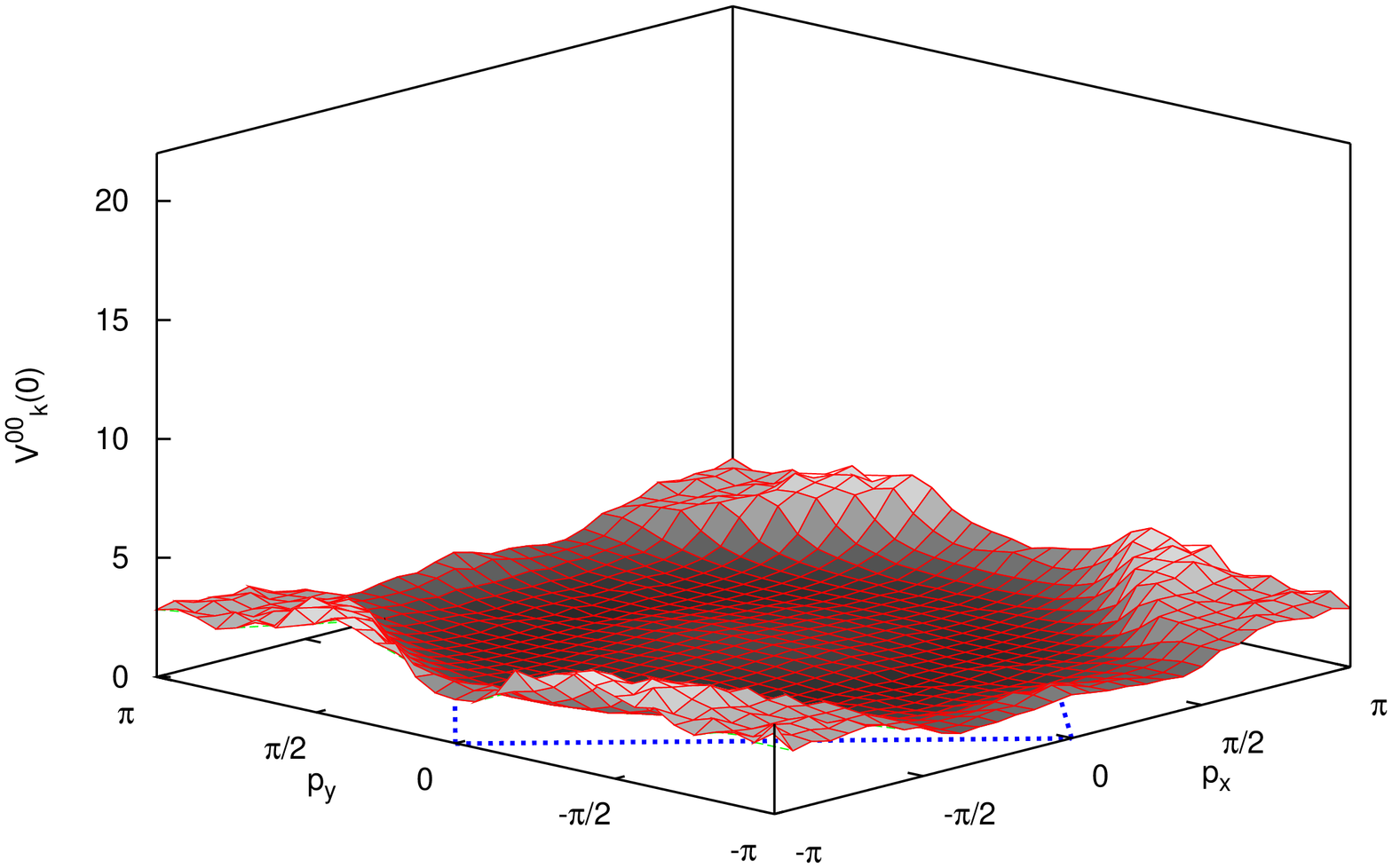}
\par\end{center}%
\end{minipage}

\begin{minipage}[t][1\totalheight]{1\columnwidth}%
\begin{flushleft}
b)
\par\end{flushleft}

\begin{center}
\includegraphics[angle=270,width=1\columnwidth]{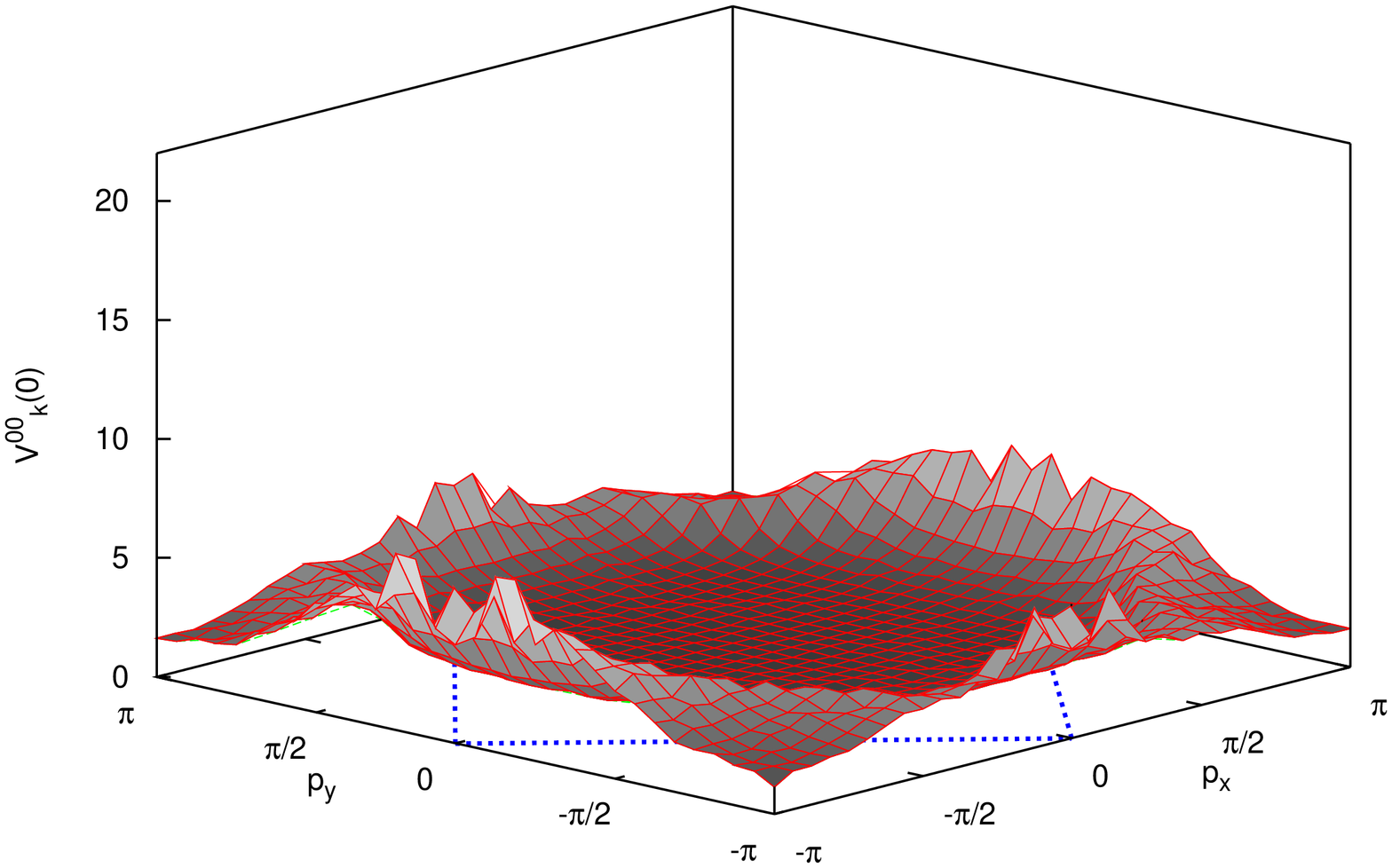}
\par\end{center}%
\end{minipage}

\begin{minipage}[t][1\totalheight]{1\columnwidth}%
\begin{flushleft}
c)
\par\end{flushleft}

\begin{center}
\includegraphics[angle=270,width=1\columnwidth]{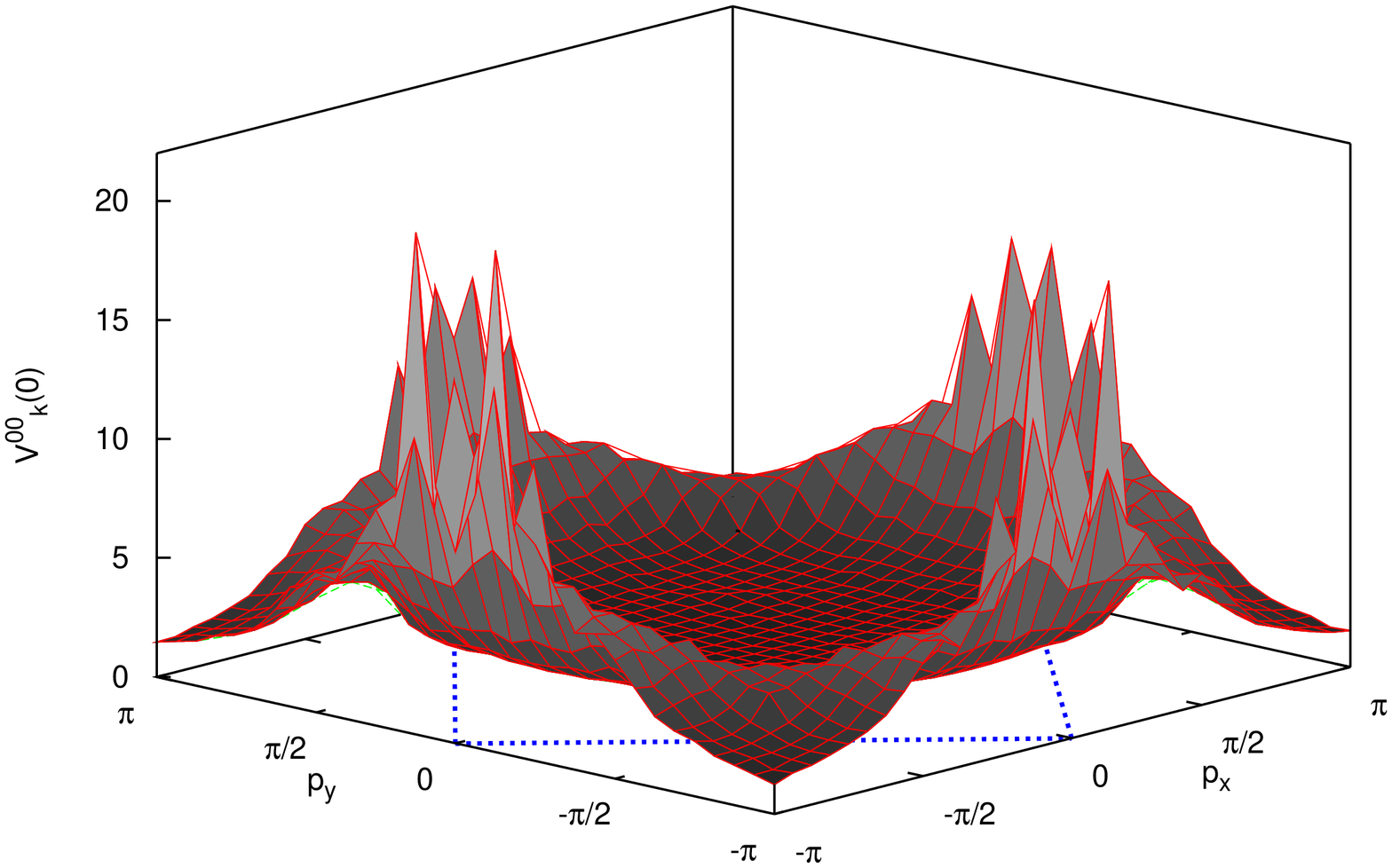}
\par\end{center}%
\end{minipage}

\caption{(Color online) $a_{3}a_{1},a_{2}a_{4}=\left(0,0\right)$ component
of the pairing interaction $\Gamma_{ph}^{a_{3}a_{1},a_{2}a_{4}}\left(\mathbf{p},\omega_{n}=0\right)$,
Eq. (\ref{ns2}), for $J=0.0\mathrm{eV}$ (a), $0.25\mathrm{eV}$(b),
and $0.50\mathrm{eV}$ (c). Pairing interaction becomes increasingly
commensurate as the Hund's coupling $J$ increases. }

\label{fig:Vplot}

\end{figure}

In Fig. \ref{fig:Vplot} we show the momentum dependence of the $a_{i}=0$
component of the effective interaction $\Gamma_{ph}^{a_{1}a_{3},a_{4}a_{2}}\left(\mathbf{p},\omega_{n}=0\right)$.
This is one of the dominating components. Other matrix elements of
$\widetilde{\Gamma}_{ph}\left(q\right)$ have a similar momentum dependence.
Finally $\widetilde{\Gamma}_{ph}\left(q\right)$ and the particle
particle interaction $\widetilde{\Gamma}_{pp}\left(q\right)$ behave
very similar. The three panels show the effective interaction mediated
by collective spin and charge fluctuations for three different values
of the Hund's coupling $J_{H}$. We clearly see that the effect of
$J_{H}$ is two-fold. On the one hand, larger values of the exchange
coupling lead to an increase of the Stoner enhancement in $\Gamma_{pp}$
and $\Gamma_{ph}$. In addition, the effective interaction becomes
increasingly more commensurate as $J_{H}$ increases. The strong peaks
close to $\mathbf{p=}\left(\pm\pi/a,0\right)$ and $\mathbf{p=}\left(0,\pm\pi/a\right)$
are consistent with the observed Bragg peaks for the magnetic ordering
in the undoped parent compounds.\cite{DeLaCruz08}

\begin{figure}
\begin{centering}
\includegraphics[width=1\columnwidth]{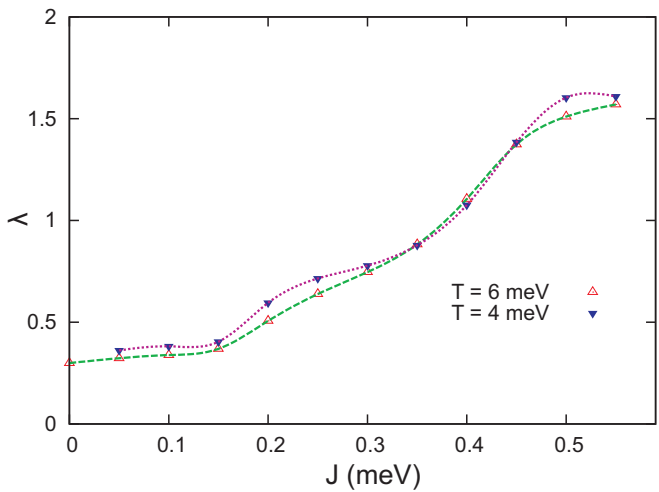}
\par\end{centering}

\caption{(Color online) Largest eigenvalue $\lambda$ for the linearized version
of Eq. (\ref{lingap}) with $\Gamma^{a_{3}a_{1},a_{2}a_{4}}\left(\mathbf{p},\omega_{n}=0\right)$
as a function of Hund's coupling $J$ for $T=0.004$eV and $0.006$eV. }

\label{fig:eigen} 
\end{figure}

In Fig. \ref{fig:eigen} we show the variation of the largest eigenvalue
$\lambda$ as function of the exchange and Hund's coupling $J$ for
two temperatures $T=0.004\mathrm{eV}\simeq46\mathrm{K}$ and $T=0.006\mathrm{eV}\simeq70\mathrm{K}$.
The enhancement of the effective pairing interaction, discussed in
Fig.3, is the primary reason for the enhancement of the pairing strength
and, in turn, of the leading eigenvalue $\lambda$. We also find that
$\lambda=1$ for $J\simeq0.4\mathrm{eV}$, which would correspond
to a critical temperature $T_{c}\simeq70\mathrm{K}$. While the above
mentioned static approximation tends to overestimate $T_{c}$, these
results demonstrate that experimentally relevant $T_{c}$ values are
clearly possible within the spin fluctuation approach.

\begin{figure}

\begin{centering}
\includegraphics[width=0.9\columnwidth]{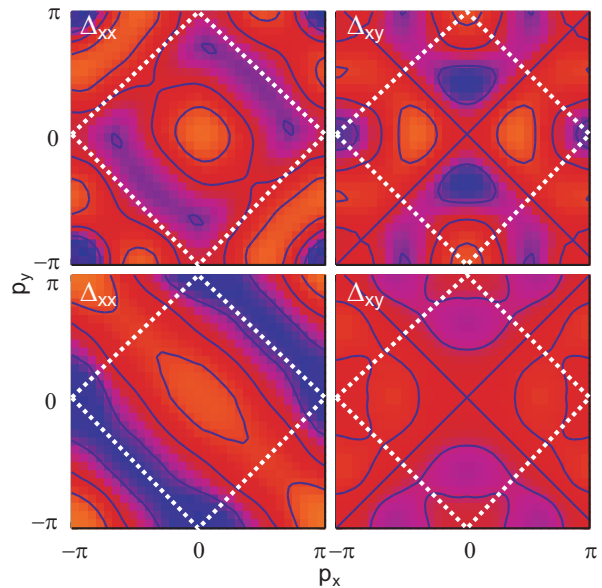}
\par\end{centering}

\caption{(Color online) Momentum dependence of $\Delta_{xx}$ and $\Delta_{xy}$
determined from the eigenvector corresponding to the leading eigenvalue
of the linearized gap equation at $T=0.006$eV for Hund's coupling
$J=0.05$eV (top) and $J=0.25$eV (bottom). Gaps of the two $xz$
orbitals are shown in an extended zone scheme. White diamonds indicate
the Brillouin zone boundary. Red (light) and blue (dark) regions correspond
to opposite signs of the gap. }

\label{fig:gap_orbit} 
\end{figure}

In Fig. \ref{fig:gap_orbit} we show the momentum dependence of $\Delta_{xx}\left(\mathbf{p}\right)$
and $\Delta_{xy}\left(\mathbf{p}\right)$ as determined from the leading
eigenvector of the linearized gap equation at $T=0.006\mathrm{eV}$.
The indicated diamond corresponds to the Brillouin zone boundary,
i.e. we plot the gap of the two $xz$ orbitals within the unit cell
in an extended zone scheme. The fact that both gap functions are of
comparable magnitude reflects the fact that Cooper pairs are formed
out of electrons in the same and in different $d$-states. The symmetry
of the gap function is $s$-wave, i.e. it is invariant with respect
to the point group operations of the Hamiltonian. Simultaneous rotation
of momenta $\mathbf{p}$ and orbitals by $\pi/2$ yields $\Delta_{xx}\left(p_{x},p_{y}\right)\rightarrow\Delta_{yy}\left(p_{y},-p_{x}\right)$
and $\Delta_{xy}\left(p_{x},p_{y}\right)\rightarrow-\Delta_{yx}\left(p_{y},-p_{x}\right)$.
The latter expression explains the sign change of $\Delta_{xy}\left(\mathbf{p}\right)$
upon rotation. It is a consequence of the $s$-wave symmetry in a
two orbital problem where the $xz$ and $yz$ orbitals transform like
the two dimensional coordinates. The fact that the diagonal gap $\Delta_{xx}\left(\mathbf{p}\right)$
differs for momenta pointing along the two diagonals of the Brillouin
zone is a consequence of the fact that the wave functions for the
$xz$ and $yz$ orbitals are different, see Ref.\cite{Qui08}. Changing
the value of the exchange interaction does not change the symmetry
of the gap function. However, it significantly affects the momentum
dependence of $\Delta_{a_{1}a_{2}}\left(\mathbf{p}\right)$. As mentioned,
the pairing interaction for small $J_{H}$ is incommensurate with
peaks rather far away from the ordering vector $\left(\pi/a,\pi/a\right)$
of the antiferromagnetic state in undoped systems at ambient pressure.
On the other hand, for $J=0.25\mathrm{eV}$, the dynamic magnetic
susceptibility and the pairing interaction $\Gamma_{pp}\left(\mathbf{p}\right)$
are peaked very close to $\left(\pi/a,\pi/a\right)$. A commensurate
pairing interaction can more efficiently change the sign of the gap
function in momentum and orbital space, while incommensurations tend
to frustrate an optimally shaped pairing gap. This leads to the more
complex pairing state for small $J$. %

Finally we determine the consequences of this gap function and analyze
the gap anisotropy on the Fermi surface. From the self energy $\Sigma_{\mathbf{k}}^{\alpha\beta}\left(i\omega_{n}\right)$
we determine the quasiparticle energies $E_{\mathbf{p}}^{\ast\alpha\beta}=E_{\mathbf{p}}^{\alpha\beta}+\Sigma_{\mathbf{k}}^{\alpha\beta}\left(0\right)-\mu\delta_{\alpha\beta}$
and construct the quasiparticle energies of the superconducting state
from the eigenvalues of \begin{equation}
\widehat{h}_{\mathbf{p}}=\left(\begin{array}{cc}
\widehat{E}_{\mathbf{p}}^{\ast} & \widehat{\Delta}_{\mathbf{p}}\\
\widehat{\Delta}_{\mathbf{p}} & -\widehat{E}_{-\mathbf{p}}^{\ast}\end{array}\right).\end{equation}
\begin{figure}
\begin{minipage}[t][1\totalheight]{1\columnwidth}%
\begin{flushleft}
a)
\par\end{flushleft}

\begin{center}
\includegraphics[scale=0.5,angle=270]{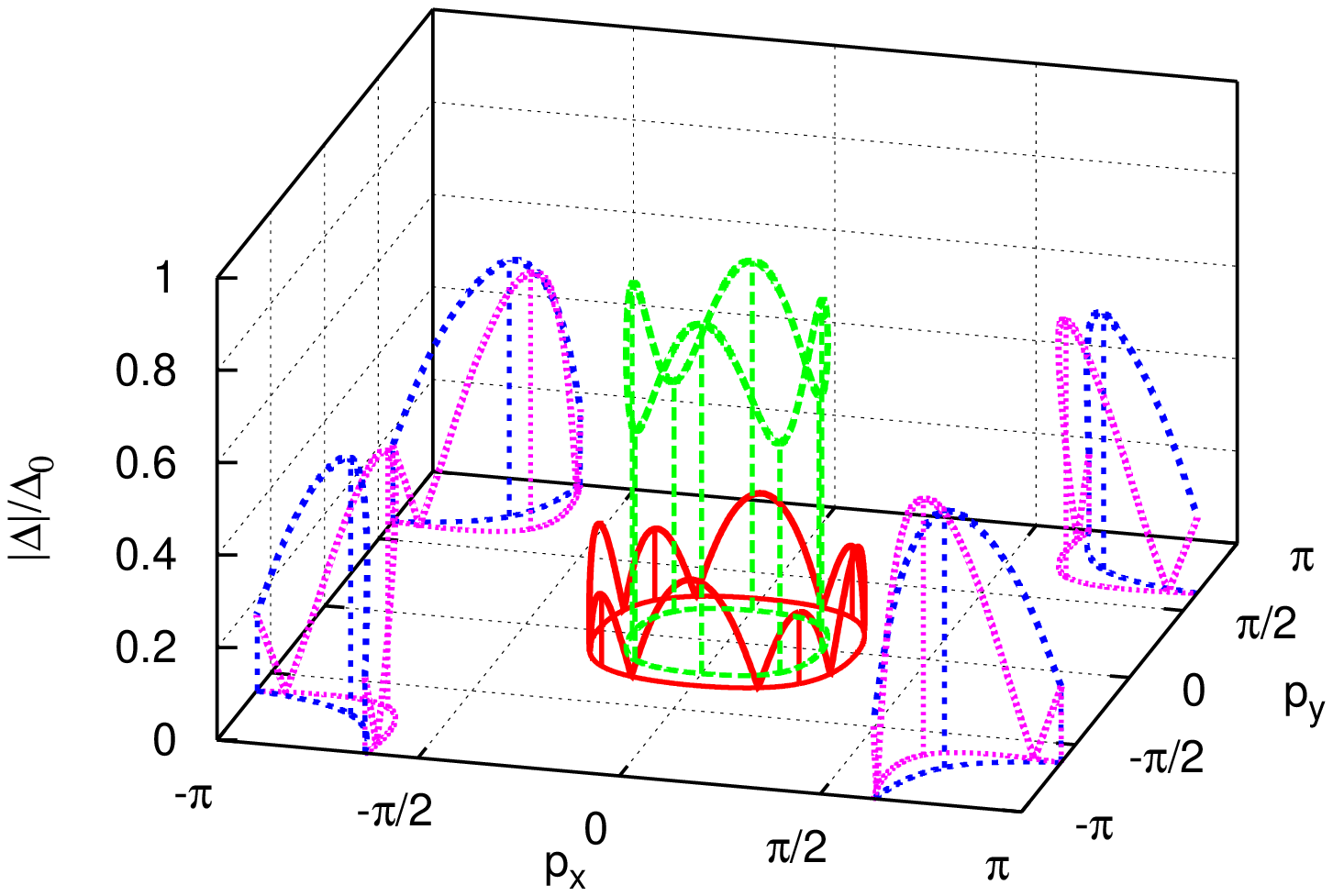} 
\par\end{center}%
\end{minipage}

\begin{minipage}[t][1\totalheight]{1\columnwidth}%
\begin{flushleft}
b)
\par\end{flushleft}

\begin{center}
\includegraphics[scale=0.5,angle=270]{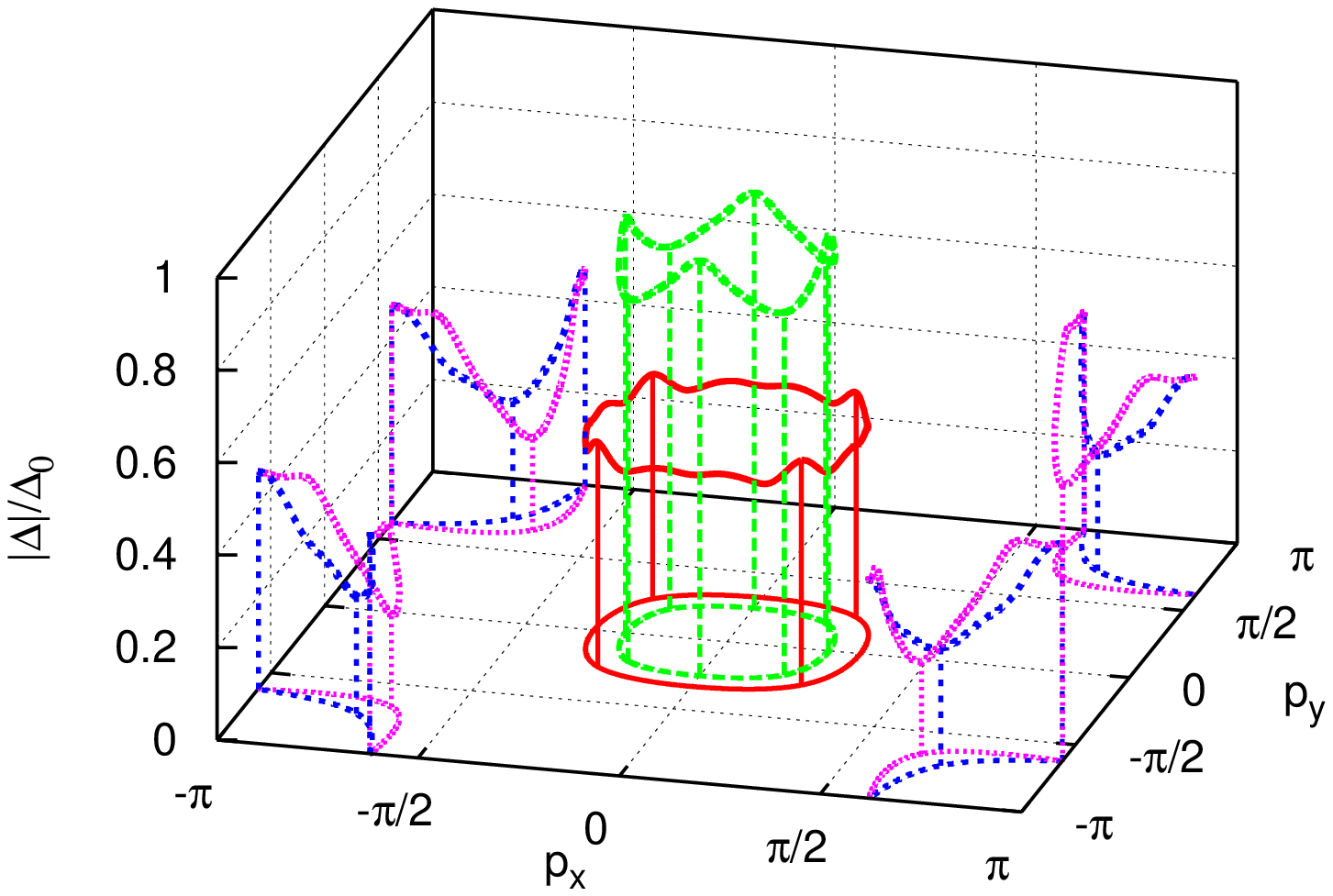} 
\par\end{center}%
\end{minipage}

\begin{minipage}[t][1\totalheight]{1\columnwidth}%
\begin{flushleft}
c)
\par\end{flushleft}

\begin{center}
\includegraphics[scale=0.5,angle=270]{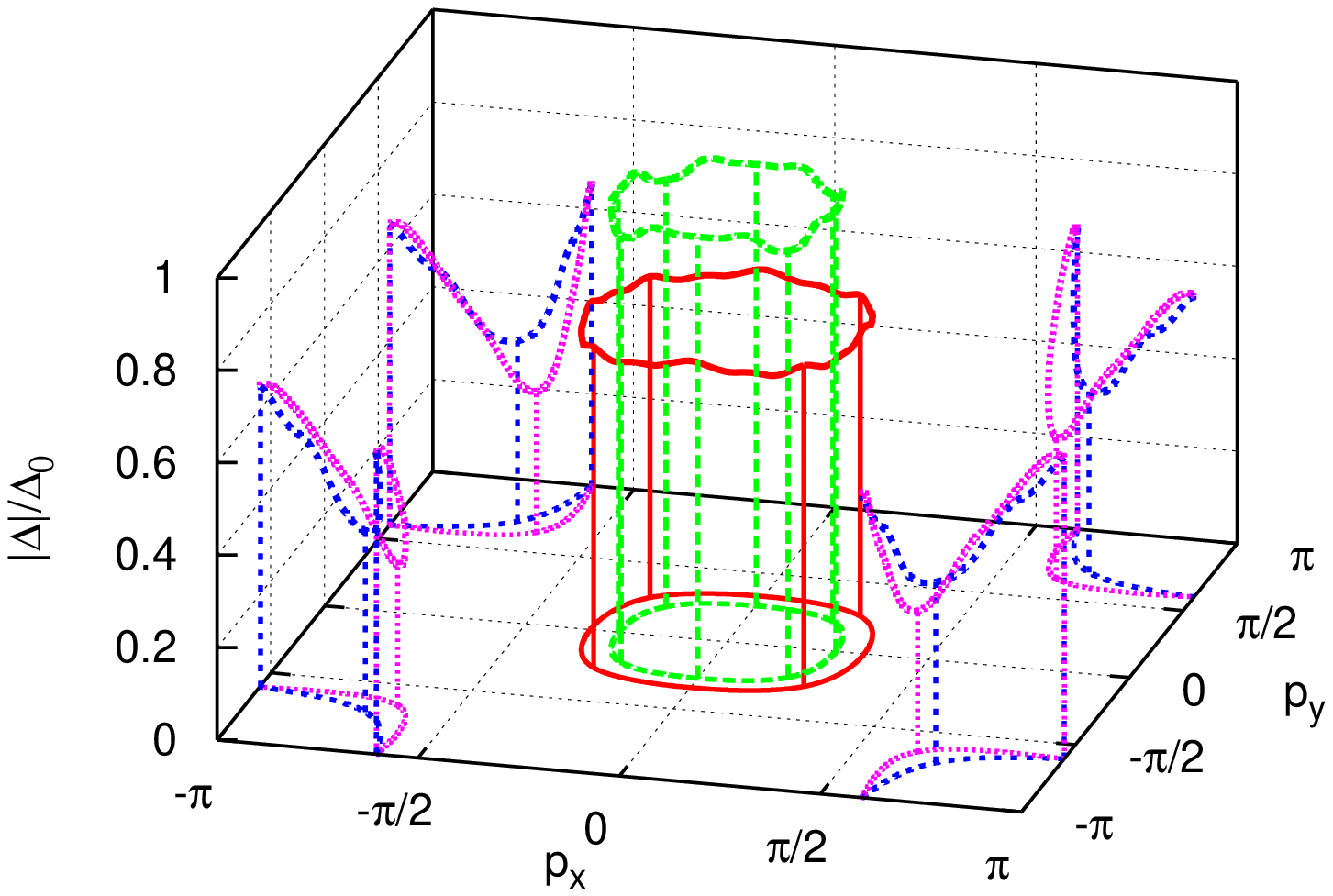} 
\par\end{center}%
\end{minipage}

\caption{(Color online) Amplitudes of the gaps along four sheets of the Fermi
surface for $J=0.0$eV (a), $0.25$eV (b), and $0.50$eV (c). While
linearized gap equation cannot define the absolute amplitude $\Delta_{0}$,
the relative gap amplitudes are properly defined.}

\label{fig:gap_band} 
\end{figure}

In Fig. \ref{fig:gap_band} we plot the magnitude of the gap along
the various sheets of the Fermi surface. The Fermi surface is constructed
from the minima of the magnitude of the eigenvalues of $\widehat{h}_{\mathbf{p}}$.
As shown in Fig. \ref{fig:gap_band}a, we find that in case of a small
$J$ the pairing interaction is more incommensurate and the gap vanishes
on line nodes on the Fermi surface. However, for larger $J$ values
we only find moderately anisotropic gap amplitudes on the Fermi surface,
see Fig. \ref{fig:gap_band}b,c. The gap amplitude on the inner Fermi
surface sheet around $\Gamma$ is significantly larger than the gap
on the outer sheet, in agreement with recent ARPES experiments.\cite{Zhao08,Ding08}
This is a consequence of the fact that $\Delta_{xx}\left(\mathbf{p}\right)$
and $\Delta_{yy}\left(\mathbf{p}\right)$ change sign close to the
Brillouin zone center. The gap of the Fermi surface sheets centered
around $M$ are considerably more anisotropic and could be responsible
for the observation of anisotropic gaps.\cite{Nakai08,Matano08,Grafe08,Mukuda08,Martin08}
In general, experiments that are sensitive to the minimum of the gap
should therefore find much smaller typical gap values and more anisotropic
gaps than measurements that are more sensitive to the largest gap
values.

Our calculation yields a fully gapped Fermi surface in the case where
the pairing interaction is close to being commensurate. In this case
the nodes of the gap are located between different Fermi surface sheets,
explaining the dramatic change in the amplitude of the gap as one
gets closer to the nodal lines (see Fig.6). The position of these
nodes is not fixed by symmetry and, as is seen in case for more incommensurate
pairing interactions, can in principle touch the Fermi surface (see
Fig.6 a). It is therefore an interesting question to ask what happens
if one includes electron-electron overlap between different FeAs layers.
This seems particularly relevant for the 122 materials where the outer
sheet of the Fermi surface around $\Gamma=\left(0,0\right)$ increases
its radius for increasing $k_{z}$.\cite{Liu08} If the pairing interaction
is predominantly two dimensional, and determined by those Fermi surface
sheets that are less dispersive in the $z$-direction, we expect that
the position of the nodes is only weakly affected by the dispersion
along $k_{z}$. It is therefore easily possible that at least one
Fermi surface sheet touches the nodal plane for larger $k_{z}$ values.
The intersection between nodal plane and Fermi surface would then
yield a nodal line on the Fermi surface. This implies that one can
easily explain fully gapped pairing states and states with line nodes
with same pairing symmetry ($s^{\pm}$) and due to the same pairing
mechanism. Note, this is impossible for a $d$-wave pairing state,
which will always yield line nodes given that the Fermi surface around
the $\Gamma$ point is closed. It is also impossible within a conventional
$s$-wave pairing state where the sign of the gap is the same everywhere.
Thus, seemingly conflicting observations in different FeAs-based systems
do not necessarily imply that there are several distinct pairing mechanism
at work.

In summary, we determined the anisotropy of the spin fluctuation induced
pairing gap on the Fermi surface of the FeAs based superconductors.
For realistic parameters we find a fully gapped state, while a measurable
anisotropy remains for some Fermi surface sheets. This may explain
the conflicting observations for the presence of gap nodes obtained
in NMR, penetration depth and ARPES experiments. It does explain the
variation of the gap on distinct sheets of the Fermi surface, as seen
in ARPES experiments.\cite{Ding08} More generally, our results demonstrate
that a fully gapped superconducting state is fully consistent with
an unconventional pairing mechanism.

We are grateful to S. L. Bud'ko, P. C. Canfield, A. V. Chubukov, V.
Cvetkovi\'{c}, A. Kaminski, I. Mazin, R. Prozorov, and J. Zhang for
helpful discussions. We express special thanks for continued interest and inspiration to B. N. Harmon. This research was supported by the Ames Laboratory,
operated for the U.S. Department of Energy by Iowa State University
under Contract No. DE-AC02-07CH11358.

\end{document}